\documentclass[AMA,STIX1COL]{WileyASNA-v1}
\usepackage{moreverb}
\usepackage{graphicx}	% Including figure files
\usepackage{amsmath}	% Advanced maths commands
\usepackage{amssymb}	% Extra maths symbols

\newcommand\BibTeX{{\rmfamily B\kern-.05em \textsc{i\kern-.025em b}\kern-.08em
T\kern-.1667em\lower.7ex\hbox{E}\kern-.125emX}}

\articletype{Original Paper}%
\received{<day> <Month>, <year>}
\revised{<day> <Month>, <year>}
\accepted{<day> <Month>, <year>}

\begin{document}

\title{Identifying interesting planetary systems for future X-ray observations}

\author[1,2]{G. Foster*}

\author[1,2]{K. Poppenhaeger}

\authormark{Foster {\&}  Poppenhaeger}

\address[1]{\orgname{Leibniz Institute for Astrophysics Potsdam (AIP)}, \orgaddress{An der Sternwarte 16, 14482 Potsdam, \country{Germany}}}

\address[2]{\orgdiv{Institute for Physics and Astronomy}, \orgname{Potsdam University}, \orgaddress{Karl-Liebknecht-Str. 24/25, 14476 Potsdam-Golm, \country{Germany}}}

\corres{*G. Foster
\email{gfoster@aip.de}}

\presentaddress{Leibniz Institute for Astrophysics Potsdam (AIP), An der Sternwarte 16, 14482 Potsdam, Germany}

\abstract[Abstract]{X-ray observations of star-planet systems are important to grow our understanding of exoplanets; these observation allow for studies of photoevaporation of the exoplanetary atmosphere, and in some cases even estimations of the size of the outer planetary atmosphere. The German-Russian eROSITA instrument onboard the SRG (Spectrum Roentgen Gamma) mission is performing the first all-sky X-ray survey since the 1990s, and provides X-ray fluxes and spectra of exoplanet host stars over a much larger volume than was accessible before. Using new eROSITA data as well as archival data from XMM-Newton, Chandra and ROSAT we estimate mass loss rates of exoplanets  under an energy-limited escape scenario, and identify several exoplanets with strong X-ray irradiation and expected mass-loss that are amenable to follow-up observations at other wavelengths. We model sample spectra using a toy model of an exoplanetary atmosphere to predict what exoplanet transit observations with future X-ray missions such as Athena will look like, and estimate the observable X-ray transmission spectrum for a typical Hot Jupiter-type exoplanet. }


%\jnlcitation{\cname{%
%\author{Foster G.}, and
%\author{K. Poppenhaeger}, (\cyear{2021}), 
%\ctitle{Identifying interesting planetary systems for future X-ray observations}, \cjournal{XXX}, \cvol{XXX}.}

\maketitle

\section{Introduction}\label{sec1}

Since the first exoplanet detected around a main-sequence star, the hot Jupiter 51 Peg b \citep{Mayor1995}, exoplanets have continued to be detected around a diverse array of host stars. \cite{Mulders2018} suggest that between 45\% and 100\% of stars have at least one planetary companion and, in addition, systems without planets interior to the orbits of Mercury and Venus are rare. This means our solar system is likely abnormal in its distance to the host star, with most planets being much closer in to their host star than we are to our Sun. Due to their close orbits, many exoplanets are subject to strong stellar irradiation. This intense radiation can cause the gas in the exoplanet atmospheres to expand and be lifted out of the gravitational well of the planet, causing the atmosphere to evaporate \citep{Owen2012}. In general, planet evaporation is thought to be driven by soft X-rays and extreme-ultraviolet (EUV) radiation received by the planet from its host star \citep{Yelle2004, Murray-Clay2009}. Whilst this EUV component is not directly observable with currently operational space observatories, the X-ray components may be observed from a number of different instruments.

As the driver of exoplanetary evaporation, high-energy irradiation is one of the most important input quantities of exoplanet evaporation rates. Thus, we are particularly interested in the X-ray and EUV irradiation of exoplanets. This X-ray irradiation of the planet can be calculated from the X-ray observations of their host stars. In turn, the stellar EUV flux can be estimated from the stellar X-ray emission and the UV part of the stellar spectrum \citep{Sanz-Forcada2011, France2013}.

We present here a characterisation of the high-energy environment of known exoplanets and estimate their mass loss rates using new X-ray data from eROSITA as well as archival data from other missions; in addition, we identify a number of systems that are interesting for follow-up observations with current or future X-ray missions, for example Athena.

\subsection{eROSITA}
eROSITA is the primary instrument on board the Spectrum Roentgen Gamma (SRG) mission \citep{Sunyaev2021, Brunner2021}. The instrument consists of seven mirror modules and is sensitive to the soft X-ray regime  \citep{Sunyaev2021}. Launched in July 2019, eROSITA is producing an all-sky survey over four years called the eROSITA All-Sky Survey (eRASS) \citep{Predehl2021}. This survey will consist of eight independent X-ray maps, one produced every six months over the four years of continuously scanning the sky \citep{Sunyaev2021}. 

We use data from the intermediate consortium-wide data release of  the first and second full-sky surveys performed by eROSITA (eRASS1 and eRASS2 respectively) proprietary to the German eROSITA collaboration (i.e.\ with a galactic longitude higher than 180$^\circ$). The eROSITA Science Analysis Software System (eSASS) (see \cite{Brunner2021}) was used to process the raw data, after which parameters such as positions and count rates in three energy bands, 0.2-0.6 keV, 0.6-2.3 keV, were listed. We focus on these extracted energy bands of the eRASS catalogue for this project.

\section{Data Analysis}

To gather the current information of X-ray irradiation of exoplanets as fully as possible, we use new eROSITA data along with archival data from XMM-Newton, Chandra and ROSAT.
We use the NASA Exoplanet Archive catalogue downloaded on March 26, 2021\footnote{\url{https://exoplanetarchive.ipac.caltech.edu}} as the basis for cross matching with the individual X-ray catalogues. We exclude any exoplanets detected by the microlensing method due to their stellar distances having uncertainties of about 50\%, which would propagate into our final exoplanetary mass-loss rates as very large uncertainties. We also discard the exoplanet around the cataclysmic variable HU Aqr, as the planet has been shown to be spurious \citep{Schwope2014, Bours2014, Gozdziewski2015}. 

Pre-eROSITA, the total number of of these planet-hosting stars that had been detected in the X-ray detected was 169 and through eRASS1 and eRASS2 this number increased by 74 to 243. We can expect this increase to be  roughly double with the data from the Russian half of the eROSITA data as well as increasing further with the upcoming three years of eRASS surveys.

\subsection{Flux Comparisons}

The fluxes reported in the X-ray catalogues used were each provided in slightly different energy bands. Therefore, before any further analysis could be done we opted to convert all fluxes into the commonly used 0.2-2.0 keV energy band. The following conversions were used for each telescope.

\textit{eROSITA}: The eRASS stellar fluxes were calculated by applying the relative conversion factor to the power-law-derived fluxes from the intermediate eRASS catalogues. This factor was found to be  $F_\mathrm{X,\,coronal} = 0.85 F_\mathrm{X,\,powerlaw}$ (see \cite{Foster2021}).

\textit{ROSAT}: Assuming a typical coronal temperature of $kT=0.3$\ keV, we to transform the ROSAT fluxes from the 0.1-2.4 keV band into the 0.2-2 keV band, using the WebPIMMS\footnote{\url{https://heasarc.gsfc.nasa.gov/cgi-bin/Tools/w3pimms/w3pimms.pl}} tool. The conversion factor is $F_\mathrm{X, 0.2-2\,keV} = 0.87 \times F_\mathrm{X, 0.1-2.4\,keV}$.

\textit{XMM-Newton}: Although the XMM-Newton catalogue already provides the 0.2-2 keV band by summing  bands 1 (0.2-0.5 keV), 2 (0.5-1.0 keV), and 3 (1.0-2.0 keV), the fluxes assume a power-law spectrum. In order to  convert this into an underlying stellar coronal model we again use WebPIMMS.  Here we find a typical correction factor of $F_\mathrm{X,\,coronal} = 0.87 F_\mathrm{X,\,powerlaw}$ for the 0.2-2 keV band for the combined signal from the EPIC cameras.

\textit{Chandra}: In its second source catalogue, Chandra lists fluxes which are not specifically model dependent. Thus we simply constructed the soft flux 0.2-2.0 keV band by combining the $u$ (0.2-0.5 keV), $s$ (0.5-1.2 keV), and $m$ (1.2-2.0 keV) bands.

After the fluxed for all data sets to cover were converted into the 0.2-2 keV range, we compared the fluxes observed for the stars which were detected by multiple missions. They were found to be in good agreement with each other (see Figure 9 in \cite{Foster2021} for the comparison between missions).

\section{Results}
\subsection{XUV Irradiation}

Incident X-ray and extreme-ultraviolet (combined, in short, to XUV) flux on a planet is thought to be the driver of the energy-limited escape process in which exoplanets are assumed to lose parts of their atmosphere \citep{Watson1981}. 

To estimate this stellar XUV fluxes we use the conversion put forward by \citet{Sanz-Forcada2011}. This approach uses stellar coronal models to calculate the extreme-ultraviolet (0.013-0.1~keV) contribution to the spectra. \citet{Sanz-Forcada2011} gives a conversion between the logs of the extreme-ultraviolet and the X-ray flux in the 0.1-2.4~keV. To change the X-ray flux in our catalogue in the 0.2-2~keV band into the required input band of 0.1-2.4\,keV band we use WebPIMMS to find a ratio of 1.15 between the two fluxes. We then calculate the EUV flux from 0.013-0.1~keV using the conversion from \citet{Sanz-Forcada2011} and add the X-ray and extreme-ultraviolet fluxes together to find the XUV flux.

\begin{figure}[thb]
\centering
\includegraphics[width=0.5\textwidth]{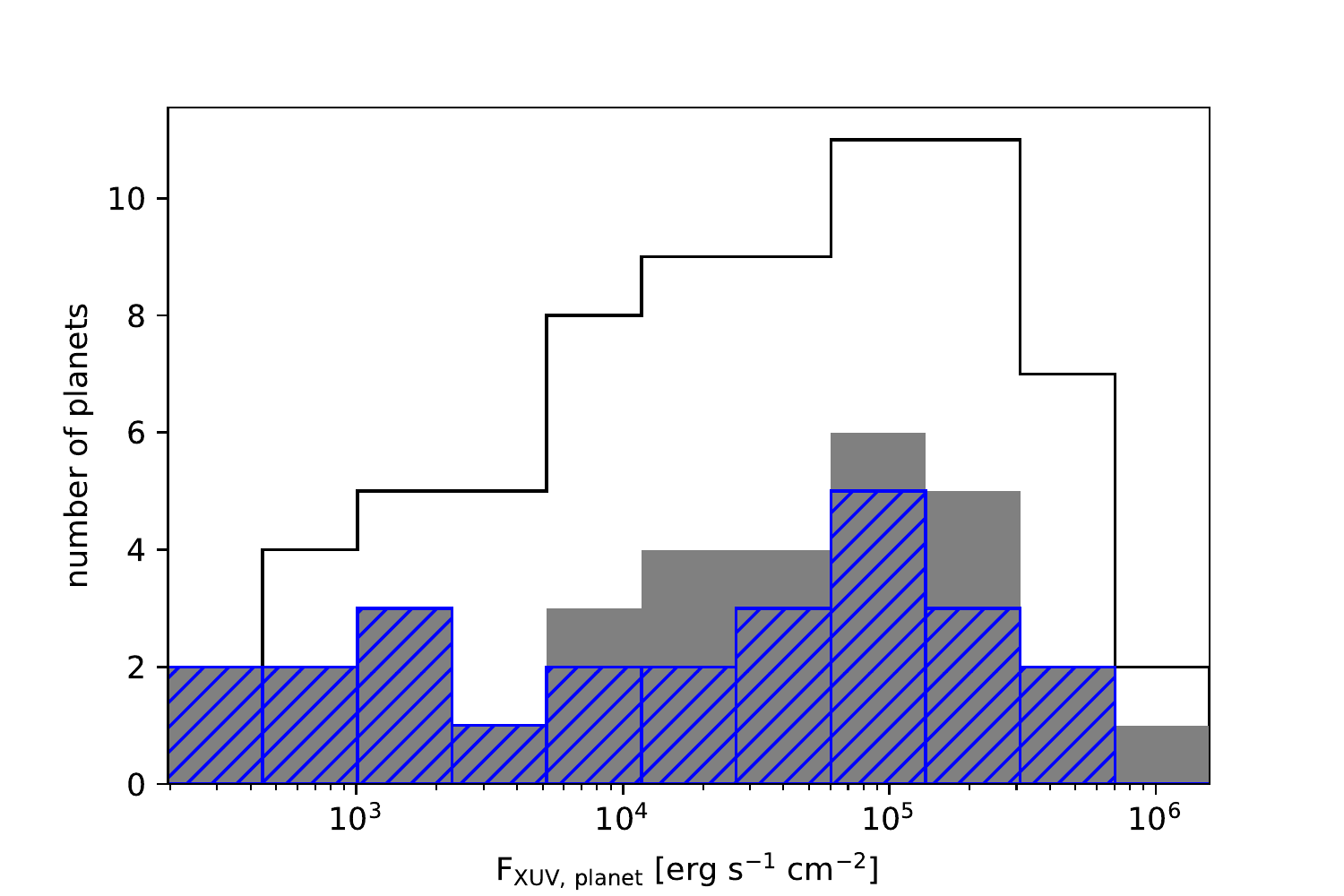}
\caption{Histogram of the XUV flux at the planetary surface of exoplanets which are transiting their X-ray detected host stars (white with black outline). All exoplanets with host stars detected with eROSITA in the first two eRASS surveys are shown in solid grey, and the planets detected from the first time in X-rays by eROSITA are shown in striped blue.}
\label{fig:fluxplanet}
\end{figure}

The XUV fluxes at the the planetary orbits shown in Fig.~\ref{fig:fluxplanet} are about five to ten times bigger than the fluxes in the X-ray 0.2-2.0~keV band on its own. The planets that we identify as particularly interesting for follow up observations are those which are highly irradiated and transiting.

Of the 59 transiting planets with X-ray detected host stars, 18 stem from new eROSITA discoveries. Comparing the XUV irradiation fluxes with that of known  evaporating exoplanets HD~189733~b and GJ~436~b, we find  a total of 16 transiting exoplanets showing irradiation levels in excess of those experienced by GJ~436~b ($8.4\times 10^{2}$\,erg\,s$^{-1}$cm$^{-2}$). Four of these exoplanets experience levels in excess of that of HD~189733~b ($8.4\times 10^{4}$\,erg\,s$^{-1}$cm$^{-2}$). This is a strong indicator that we may be able to  observe the ongoing evaporation of these exoplanets in optical and UV wavelength bands, as has been done for HD~189733~b and GJ~436~b.

\subsection{Mass-Loss Rates}
In the context of this work, we use a simple energy-limited hydrodynamic escape model based on \citet{Lopez2012} and \citet{Owen2012}, and refer to \citet{Poppenhaeger2021} for the assumptions made in our model.

\begin{figure}[thb]
\centering
\includegraphics[width=0.5\textwidth]{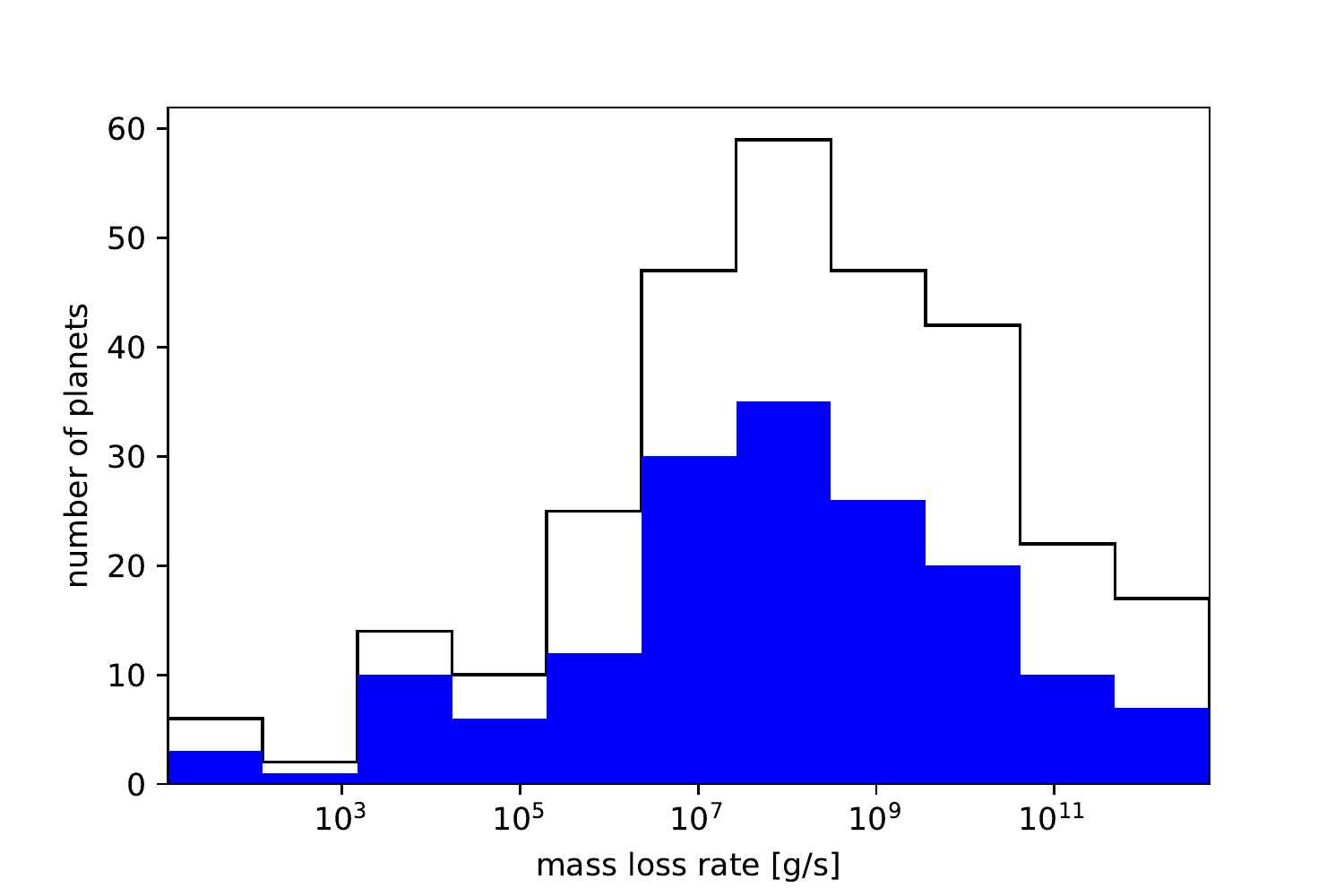}
\caption{Histogram of the estimated mass loss rates of X-ray detected exoplanets (white with black outline) and the exoplanets with host stars detected with eROSITA in the first two eRASS surveys (blue).}
\label{fig:masslosshist}
\end{figure}

In instances where either the planet mass or the planet radius were known, but not both, we estimated the planet radius from its mass, or vice versa using the mass-radius relationship given by \citet{Chen2017}. We plot the distribution of our estimate mass-losses in Fig.~\ref{fig:masslosshist}. 

The distributed mass-losses in Fig.~\ref{fig:masslosshist} omit estimates for exoplanets with radii smaller than $1.6R_\oplus$. This is based on \citet{Rogers2015}, who showed that exoplanets smaller than $1.6R_\oplus$ are unlikely to undergo any significant atmospheric mass-loss as they are likely fully rocky.

With eROSITA, we measure four expected mass-loss rates higher than those of HD~189733~b for the first time ( $1.8\times 10^{11}$\,g\,s$^{-1}$), as well as 14 new eROSITA measurements which have expected mass-loss rates higher than those of the Neptune GJ~436~b ( $1.0\times 10^{9}$\,g\,s$^{-1}$).

\section{Discussion}
\subsection{Interesting Systems}

Systems where an exoplanet is both transiting its host star and highly irradiated are most suitable for follow up observations. With X-ray observatories such as eROSITA some of these systems are being observed in the X-ray regime for the first time. Especially interesting, are binary systems which can either be resolved for the first time, or will soon be resolvable either with further eROSITA surveys or future missions such as Athena.

%\textbf{Comment by Katja: add a sentence or two about what makes an interesting system in your opinion.}

\subsubsection{DS TUC A}

DS Tuc A is G-type star and the primary stellar component of its binary system, which includes the K3 secondary, DS Tuc B \citep{Newton2019}. DS Tuc A hosts a planet, DS Tuc A b, that has an orbital period of 8.14 days \citep{Benatti2021}. The system was partially resolved by the XMM-Newton in the MOS images, with both stars in the binary appearing with similar intensities \citep{Benatti2021}. However, \cite{Benatti2021} still reports that in estimating the fluxes there is cross-contamination of order of 16\% due to the tight separation if the system.

In \cite{Foster2021} a high mass-loss rate of  $6\times 10^{11}$\,g\,s$^{-1}$ was estimated for the planet of radius 5.7\,$R_\oplus$  and mass 26.7\,$M_\oplus$. This was estimated from the high stellar X-ray flux of $3.1\times 10^{-12}$\,erg\,s$^{-1}$cm$^{-2}$, which is 50\% of the detected eRASS flux at the position of the DS Tuc binary system. 

With more observations over future eRASS surveys we hope to fully resolve the binary system, which is separated by 5$^{\prime\prime}$ \citep{Newton2019}, to make a more accurate calculation of the mass loss rate of DS Tuc A b using the actual X-ray flux of DS Tuc A.

\subsubsection{WASP-180 A}

WASP-180 A b is a transiting exoplanet whose host star has been detected in the X-ray regime for the first time with the eRASS surveys. This hot Jupiter orbits  the primary star of a visual binary, WASP-180 A \citep{Temple2019}.

This planet, which has a mass of 0.9\,$M_J$ and a radius of 1.2\,$R_J$
%286\,$M_\oplus$ and a radius of 13.9\,$R_\oplus$ 
%\textbf{(I think it would be better to give its mass and radius in Jupiter units, as it's a giant planet)}
, is highly irradiated with an XUV flux of $5.7\times 10^{5}$\,erg\,s$^{-1}$cm$^{-2}$ at the planetary surface. This gives us an estimated mass loss rate of $2\times 10^{12}$\,g\,s$^{-1}$.  This is a greater estimated mass loss rate than the that of the well known Hot Jupiter HD 189733 b (estimated to have a mass loss rate between  $10^{9}$ and $ 10^{11}$\,gs$^{-1}$ by \citet{Lecavelier2010}) making this star a good target for follow up observations at other wavelengths to potentially directly observe the mass loss of WASP-180 A b.

\subsection{Simulated Athena Spectra}
In the future we will not only have the eROSITA mission for more in-depth X-ray observations. Exoplanetary science may be furthered with the Athena's ability for spatially resolved X-ray spectroscopy and deep wide-field spectral imaging. This instrument is expected to yield a vast improvement over the capabilities of current X-ray observatories such as XMM-Newton and eROSITA.

The Advanced Telescope for High Energy Astrophysics (Athena) mission, an X-Ray observatory selected by the European Space Agency (ESA) in 2014, is currently scheduled to launch in 2031\footnote{\url{https://sci.esa.int/web/athena/-/59896-mission-summary}} \citep{Barret2020}. The observatory includes the high resolution X-Ray spectrometer called the X-Ray integral Field Unit (X-IFU) \citep{Barret2018}. This instrument will produce X-ray spectra in the 0.2 to 12 keV range with a spectral resolution of about 2.5 eV and a field of view of 5' \citep{Barret2018}. 

The X-IFU, with its higher sensitivity and spectral resolution compared to current instruments, is expected to improved our knowledge of the effects of X-ray irradiation on exoplanets immensely. With this it may be possible to measure the transmission spectrum of exoplanet atmospheres during transits for the first time, making this mission of interest for the future of exoplanetary science \citep{Barret2016}.

\begin{figure}[thb]
\centering
\includegraphics[width=0.5\textwidth]{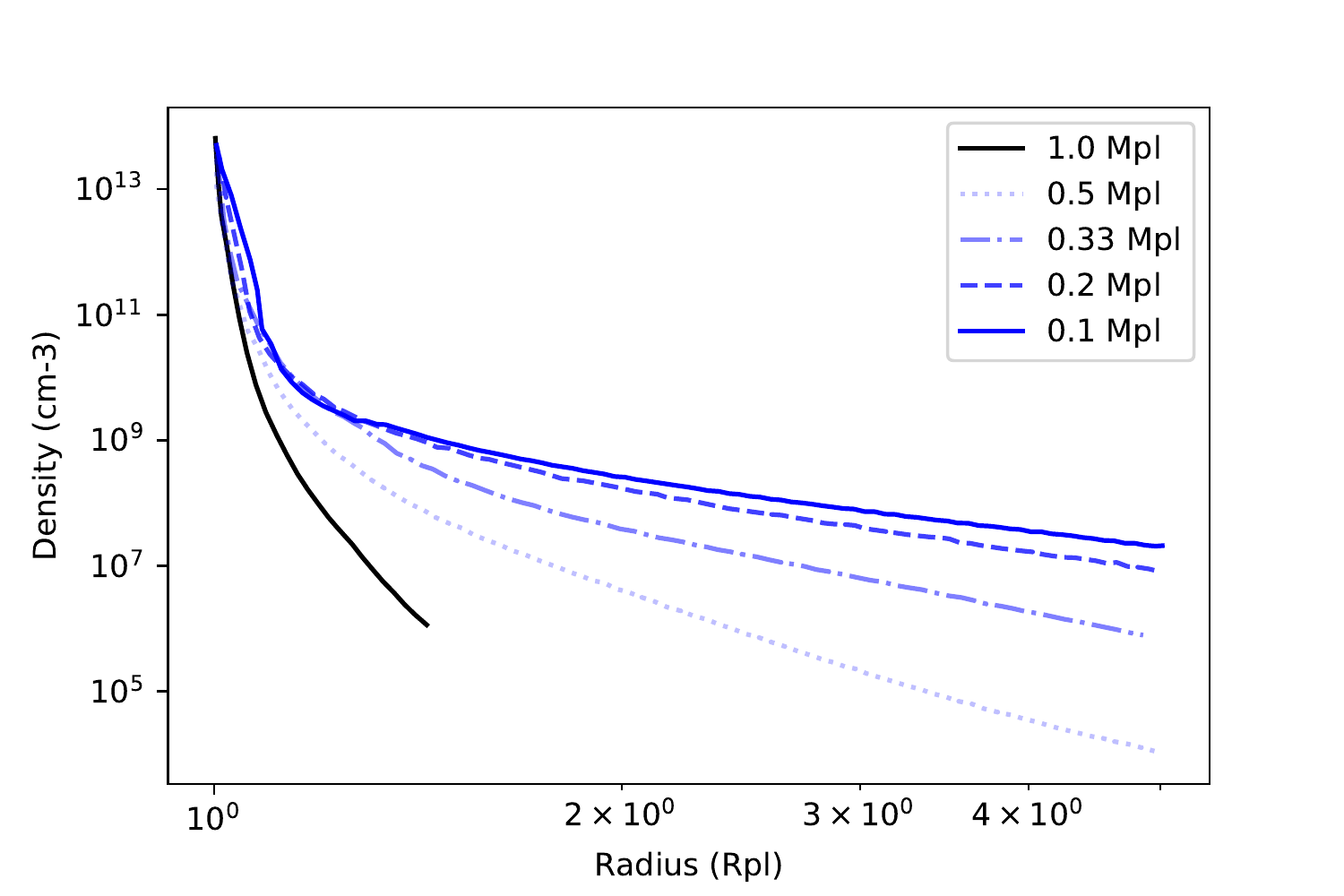}
\caption{Density of planetary atmosphere of  planets of differing masses in terms of the mass of WASP-10 b. The planets are with the same separation from their host star as that of WASP-10 b. Plot was adapted from \cite{Salz2016}.}
\label{fig:densrad}
\end{figure}

\begin{figure}[thb]
\centering
\includegraphics[width=0.5\textwidth]{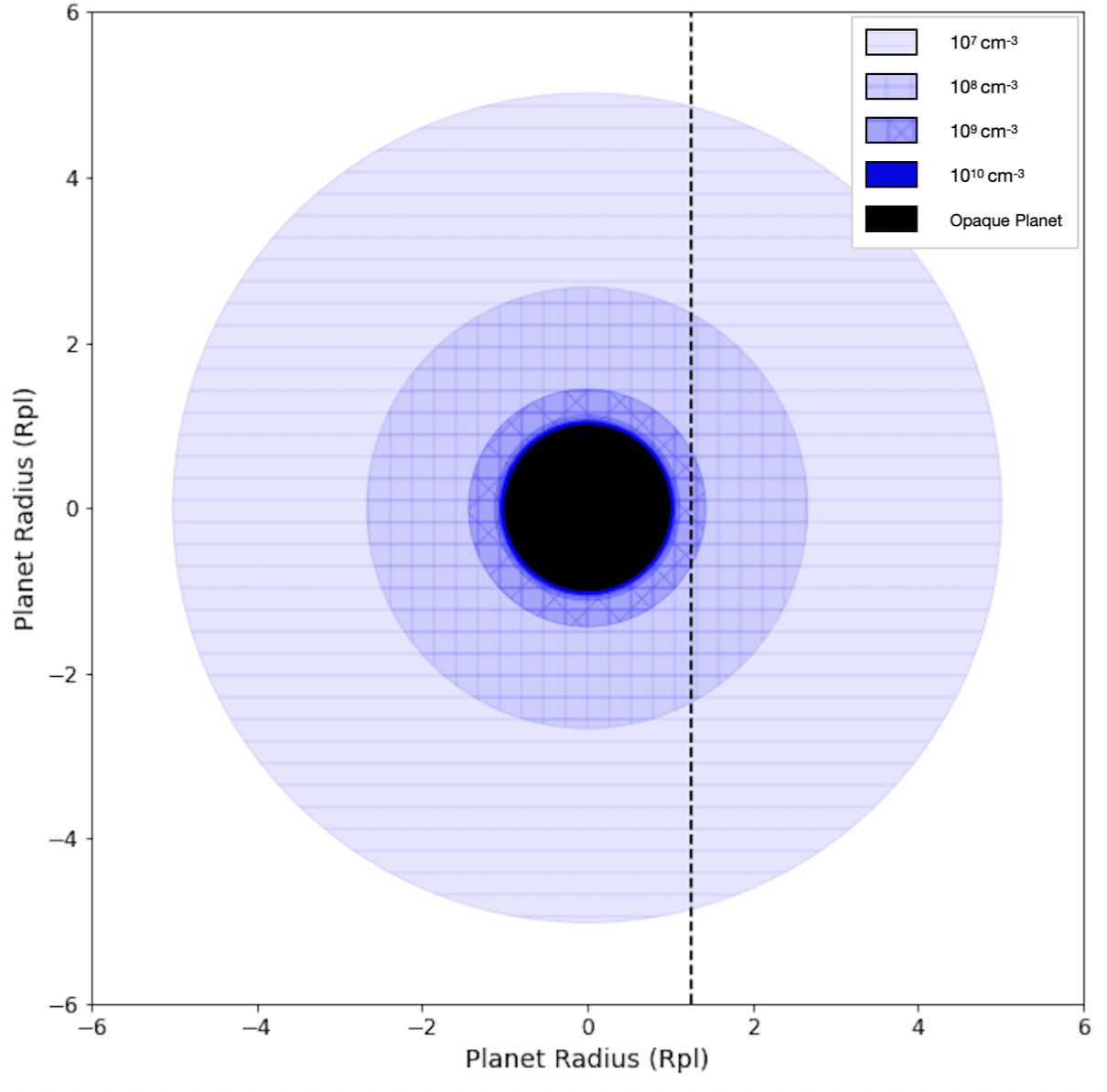}
\caption{Toy model of an exoplanet (rocky core shown by the solid black circle) and its atmosphere. The atmosphere has decreasing density as the distance from the planetary surface increases (concentric blue circles around the planet).}
\label{fig:toymodel}
\end{figure}

We use a  density radius planet model by \cite{Salz2016} as a representation of how an atmosphere changes for planets of different masses (Fig.~\ref{fig:densrad}). From this we make a toy model of an atmosphere around a planet with 10\% of the mass of WASP-10 b (Fig.~\ref{fig:toymodel}). This model simplifies a planet atmosphere into areas of descending densities as we move away from the planetary surface. We take multiple cross-sections of the planetary atmosphere to calculate the column density through each concentric ring, as well as to calculate the area of the planet of this density. An example of one cross section is shown by the dotted black line in Fig.~\ref{fig:toymodel}.

\begin{figure}[thb]
\centering
\includegraphics[width=0.5\textwidth]{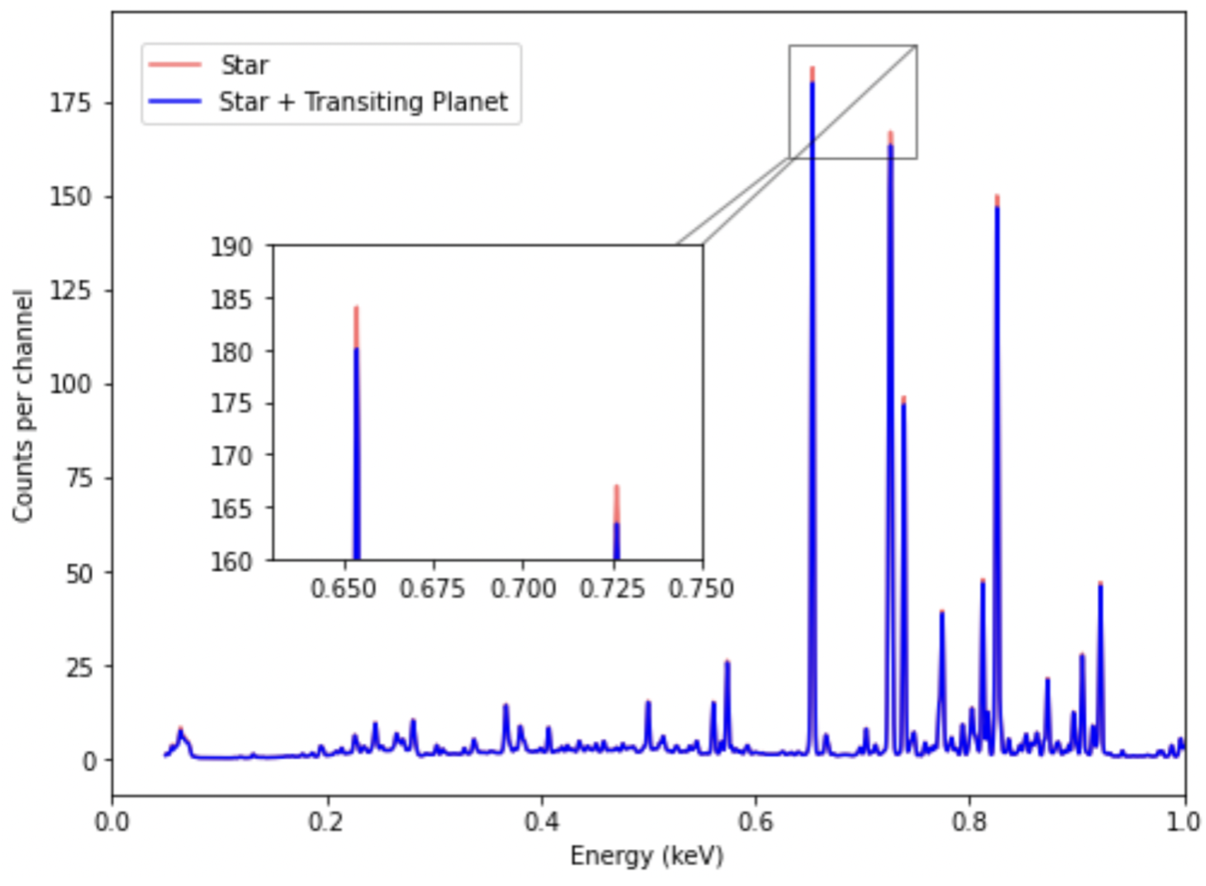}
\caption{Spectrum of a model star with a transiting planet (blue) compared to the spectrum of the same star when no planet is present (red). A magnification of the two largest peaks is shown in the subplot to highlight the change more clearly.}
\label{fig:spectra}
\end{figure}

We then simulated spectra with XSPEC version 12.11.1 \citep{Arnaud1996} using an APEC coronal model with differing absorbing columns, using multiplicative photo-electric absorption models, to simulate each ring of our toy atmosphere. We assumed solar abundances for the absorbing planetary atmosphere. We use a single temperature component with a $kT$ of 0.4 keV for each model, corresponding to a temperature of approximately 4.6 million K, and an emission measure corresponding to a bright stellar X-ray flux of about $10^{-12}$~erg/s/cm$^2$ . We show how the spectra of the star changes during a planetary transit in Fig.~\ref{fig:spectra}.

\begin{figure}[thb]
\centering
\includegraphics[width=0.5\textwidth]{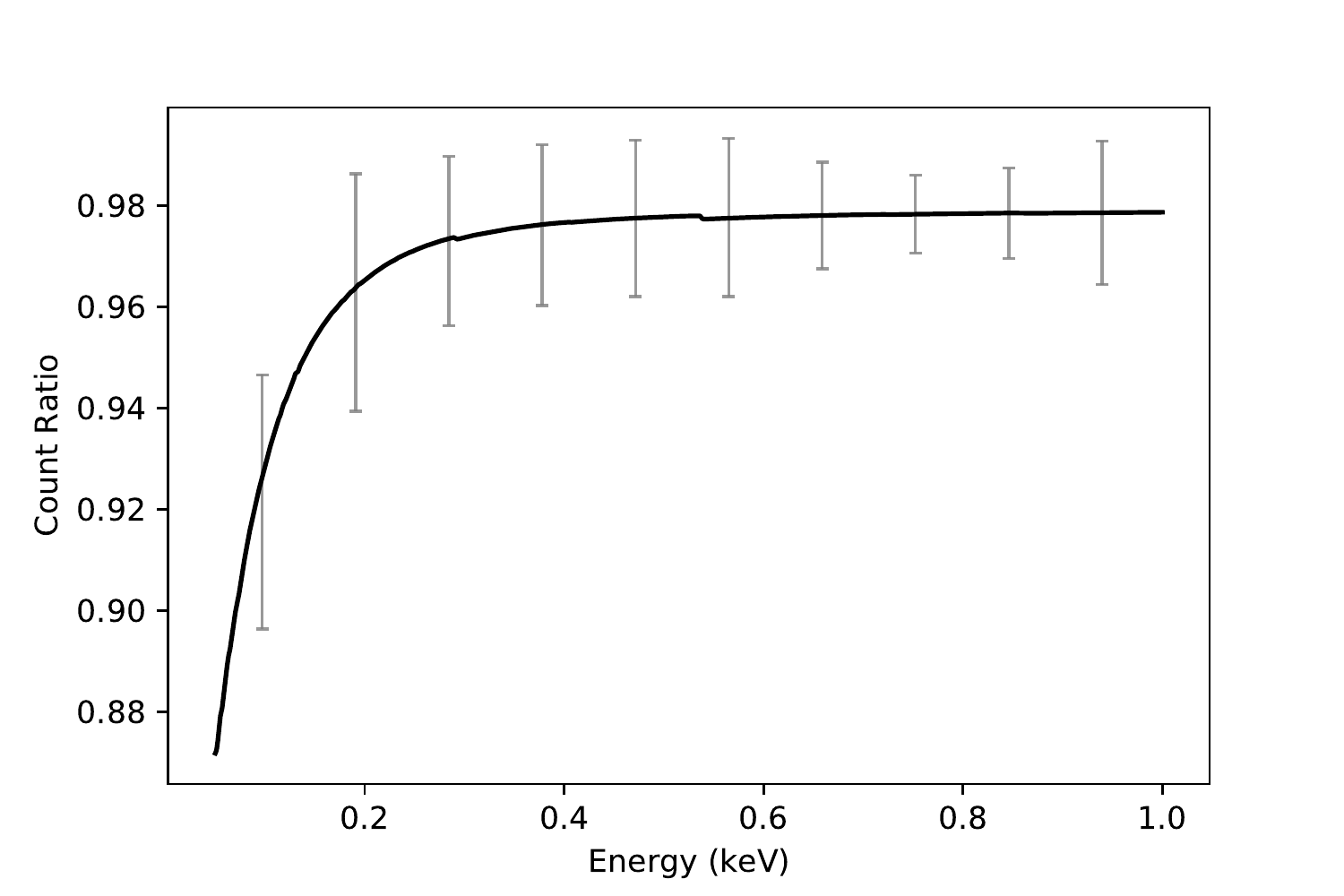}
\caption{Model transmission spectrum of a star during the planetary transit of a planet with a mass that is 10\% of the mass of WASP-10 b. Uncertainties are calculated for spectra with an accumulated exposure time of 100ks and a stellar X-ray flux of $10^{-12}$~erg/s/cm$^2$.}
\label{fig:transspec}
\end{figure}

The transmission spectra of the modelled transiting exoplanet is shown in  Fig.~\ref{fig:transspec}, as the quotient of the X-ray spectrum with the planet in transit and the planet-free spectrum, simulated for 100~ks observing time accumulated in transit and outside of transit. This corresponds to co-added observations of the order of ten observed transits for typical transit duration of short-period exoplanets.  Note the  small ionisation edges at 0.5 keV and 0.3 keV in the model, which correspond to oxygen and carbon in the atmosphere respectively \citep{Wilms2000}. However, it is the absorption at very soft X-ray energies that will likely be best observable since at those energies the X-ray radius of the planetary atmosphere increases significantly.

Although it is unlikely the oxygen absorption edge the size of the one simulated here would be detectable from the uncertainties in the spectra, it is possible that this edge could be more prominent for some transiting planets. For example with respect to solar abundances, Jupiter has an atmosphere enriched in carbon by a factor of about three \citep{Oberg2019}. Similar conditions in hot Jupiters with respect to their host stars may produce transmission spectra with more prominent absorption edges which would be more easily detectable with Athena.

\section{Conclusion}

We have presented high-energy irradiation levels and mass-loss estimates for exoplanets with X-ray observations using new eROSITA data as well as archival data from XMM-Newton, Chandra and ROSAT. Particularly interesting targets for follow-up observations, DS Tuc and WASP-180, with either eROSITA or future X-ray missions, such as Athena, were identified.

We have modelled sample spectra of stars as seen by Athena and predict what exoplanetary transit observations will look like with this observatory. We use this to estimate the observable X-ray transmission spectrum for a hot Jupiter with a mass 10\% of the mass of WASP-10 b. From this modelled transmission spectrum we identified that with future missions such as Athena the absorption of outer exoplanetary atmospheres may be identified using X-ray observations.

\section{Acknowledgments}

This work was supported by the German \textit{Leibniz-Gemeinschaft}, project number P67-2018.

This work is based on data from eROSITA, the primary instrument aboard SRG, a joint Russian-German science mission supported by the Russian Space Agency (Roskosmos), in the interests of the Russian Academy of Sciences represented by its Space Research Institute (IKI), and the Deutsches Zentrum f{\"u}r Luft und Raumfahrt (DLR). The SRG spacecraft was built by Lavochkin Association (NPOL) and its subcontractors, and is operated by NPOL with support from the Max Planck Institute for Extraterrestrial Physics (MPE). The development and construction of the eROSITA X-ray instrument was led by MPE, with contributions from the Dr. Karl Remeis Observatory Bamberg \& ECAP (FAU Erlangen-Nuernberg), the University of Hamburg Observatory, the Leibniz Institute for Astrophysics Potsdam (AIP), and the Institute for Astronomy and Astrophysics of the University of T{\"u}bingen, with the support of DLR and the Max Planck Society. The Argelander Institute for Astronomy of the University of Bonn and the Ludwig Maximilians Universit{\"a}t Munich also participated in the science preparation for eROSITA. The eROSITA data shown here were processed using the eSASS software system developed by the German eROSITA consortium. This work has made use of data from the Chandra X-ray Observatory, the ROSAT mission, and the XMM-Newton mission.

\typeout{}
\bibliography{biblio}

\begin{thebibliography}{}

\bibitem [\protect \citeauthoryear {%
{Arnaud}%
}{%
{Arnaud}%
}{%
{\protect \APACyear {1996}}%
}]{%
Arnaud1996}
\APACinsertmetastar {%
Arnaud1996}%
\begin{APACrefauthors}%
{Arnaud}, K\BPBI A.%
\end{APACrefauthors}%
\unskip\
\newblock
\APACrefYearMonthDay{1996}{{\APACmonth{01}}}{},
\newblock
{\BBOQ}\APACrefatitle {{XSPEC: The First Ten Years}} {{XSPEC: The First Ten
  Years}}.{\BBCQ}
\newblock
\BIn{} G\BPBI H.~{Jacoby}\ \BBA {} J.~{Barnes}\ (\BEDS), \APACrefbtitle
  {Astronomical Data Analysis Software and Systems V} {Astronomical Data
  Analysis Software and Systems V}\ \BVOL~101, \BPG~17.
\PrintBackRefs{\CurrentBib}

\bibitem [\protect \citeauthoryear {%
{Barret}%
\ \protect \BOthers {.}}{%
{Barret}%
\ \protect \BOthers {.}}{%
{\protect \APACyear {2020}}%
}]{%
Barret2020}
\APACinsertmetastar {%
Barret2020}%
\begin{APACrefauthors}%
{Barret}, D.%
, {Decourchelle}, A.%
, {Fabian}, A.%
, {Guainazzi}, M.%
, {Nandra}, K.%
, {Smith}, R.%
\BCBL {}\ \BBA {} {den Herder}, J\BHBI W.%
\end{APACrefauthors}%
\unskip\
\newblock
\APACrefYearMonthDay{2020}{{\APACmonth{02}}}{},
\newblock
\unskip
\newblock
\APACjournalVolNumPages{Astronomische Nachrichten}{341}{2}{224-235}.
\newblock
\begin{APACrefDOI} \doi{10.1002/asna.202023782} \end{APACrefDOI}
\PrintBackRefs{\CurrentBib}

\bibitem [\protect \citeauthoryear {%
{Barret}%
\ \protect \BOthers {.}}{%
{Barret}%
\ \protect \BOthers {.}}{%
{\protect \APACyear {2016}}%
}]{%
Barret2016}
\APACinsertmetastar {%
Barret2016}%
\begin{APACrefauthors}%
{Barret}, D.%
, {Lam Trong}, T.%
, {den Herder}, J\BHBI W.%
\ et al.\end{APACrefauthors}%
\unskip\
\newblock
\APACrefYearMonthDay{2016}{{\APACmonth{07}}}{},
\newblock
{\BBOQ}\APACrefatitle {{The Athena X-ray Integral Field Unit (X-IFU)}} {{The
  Athena X-ray Integral Field Unit (X-IFU)}}.{\BBCQ}
\newblock
\BIn{} J\BHBI W\BPBI A.~{den Herder}, T.~{Takahashi}\BCBL {}\ \BBA {}
  M.~{Bautz}\ (\BEDS), \APACrefbtitle {Space Telescopes and Instrumentation
  2016: Ultraviolet to Gamma Ray} {Space Telescopes and Instrumentation 2016:
  Ultraviolet to Gamma Ray}\ \BVOL\ 9905, \BPG~99052F.
\newblock
\begin{APACrefDOI} \doi{10.1117/12.2232432} \end{APACrefDOI}
\PrintBackRefs{\CurrentBib}

\bibitem [\protect \citeauthoryear {%
{Barret}%
\ \protect \BOthers {.}}{%
{Barret}%
\ \protect \BOthers {.}}{%
{\protect \APACyear {2018}}%
}]{%
Barret2018}
\APACinsertmetastar {%
Barret2018}%
\begin{APACrefauthors}%
{Barret}, D.%
, {Lam Trong}, T.%
, {den Herder}, J\BHBI W.%
\ et al.\end{APACrefauthors}%
\unskip\
\newblock
\APACrefYearMonthDay{2018}{{\APACmonth{07}}}{},
\newblock
{\BBOQ}\APACrefatitle {{The ATHENA X-ray Integral Field Unit (X-IFU)}} {{The
  ATHENA X-ray Integral Field Unit (X-IFU)}}.{\BBCQ}
\newblock
\BIn{} J\BHBI W\BPBI A.~{den Herder}, S.~{Nikzad}\BCBL {}\ \BBA {}
  K.~{Nakazawa}\ (\BEDS), \APACrefbtitle {Space Telescopes and Instrumentation
  2018: Ultraviolet to Gamma Ray} {Space Telescopes and Instrumentation 2018:
  Ultraviolet to Gamma Ray}\ \BVOL\ 10699, \BPG~106991G.
\newblock
\begin{APACrefDOI} \doi{10.1117/12.2312409} \end{APACrefDOI}
\PrintBackRefs{\CurrentBib}

\bibitem [\protect \citeauthoryear {%
{Benatti}%
\ \protect \BOthers {.}}{%
{Benatti}%
\ \protect \BOthers {.}}{%
{\protect \APACyear {2021}}%
}]{%
Benatti2021}
\APACinsertmetastar {%
Benatti2021}%
\begin{APACrefauthors}%
{Benatti}, S.%
, {Damasso}, M.%
, {Borsa}, F.%
\ et al.\end{APACrefauthors}%
\unskip\
\newblock
\APACrefYearMonthDay{2021}{{\APACmonth{06}}}{},
\newblock
\unskip
\newblock
\APACjournalVolNumPages{\aap}{650}{}{A66}.
\newblock
\begin{APACrefDOI} \doi{10.1051/0004-6361/202140416} \end{APACrefDOI}
\PrintBackRefs{\CurrentBib}

\bibitem [\protect \citeauthoryear {%
{Bours}%
\ \protect \BOthers {.}}{%
{Bours}%
\ \protect \BOthers {.}}{%
{\protect \APACyear {2014}}%
}]{%
Bours2014}
\APACinsertmetastar {%
Bours2014}%
\begin{APACrefauthors}%
{Bours}, M\BPBI C\BPBI P.%
, {Marsh}, T\BPBI R.%
, {Breedt}, E.%
\ et al.\end{APACrefauthors}%
\unskip\
\newblock
\APACrefYearMonthDay{2014}{{\APACmonth{12}}}{},
\newblock
\unskip
\newblock
\APACjournalVolNumPages{\mnras}{445}{2}{1924-1931}.
\newblock
\begin{APACrefDOI} \doi{10.1093/mnras/stu1879} \end{APACrefDOI}
\PrintBackRefs{\CurrentBib}

\bibitem [\protect \citeauthoryear {%
{Brunner}%
\ \protect \BOthers {.}}{%
{Brunner}%
\ \protect \BOthers {.}}{%
{\protect \APACyear {2021}}%
}]{%
Brunner2021}
\APACinsertmetastar {%
Brunner2021}%
\begin{APACrefauthors}%
{Brunner}, H.%
, {Liu}, T.%
, {Lamer}, G.%
\ et al.\end{APACrefauthors}%
\unskip\
\newblock
\APACrefYearMonthDay{2021}{{\APACmonth{06}}}{},
\newblock
\unskip
\newblock
\APACjournalVolNumPages{arXiv e-prints}{}{}{arXiv:2106.14517}.
\PrintBackRefs{\CurrentBib}

\bibitem [\protect \citeauthoryear {%
{Chen}%
\ \BBA {} {Kipping}%
}{%
{Chen}%
\ \BBA {} {Kipping}%
}{%
{\protect \APACyear {2017}}%
}]{%
Chen2017}
\APACinsertmetastar {%
Chen2017}%
\begin{APACrefauthors}%
{Chen}, J.%
\BCBT {}\ \BBA {} {Kipping}, D.%
\end{APACrefauthors}%
\unskip\
\newblock
\APACrefYearMonthDay{2017}{{\APACmonth{01}}}{},
\newblock
\unskip
\newblock
\APACjournalVolNumPages{\apj}{834}{1}{17}.
\newblock
\begin{APACrefDOI} \doi{10.3847/1538-4357/834/1/17} \end{APACrefDOI}
\PrintBackRefs{\CurrentBib}

\bibitem [\protect \citeauthoryear {%
{Foster}%
, {Poppenhaeger}%
, {Ilic}%
\BCBL {}\ \BBA {} {Schwope}%
}{%
{Foster}%
\ \protect \BOthers {.}}{%
{\protect \APACyear {2021}}%
}]{%
Foster2021}
\APACinsertmetastar {%
Foster2021}%
\begin{APACrefauthors}%
{Foster}, G.%
, {Poppenhaeger}, K.%
, {Ilic}, N.%
\BCBL {}\ \BBA {} {Schwope}, A.%
\end{APACrefauthors}%
\unskip\
\newblock
\APACrefYearMonthDay{2021}{{\APACmonth{06}}}{},
\newblock
\unskip
\newblock
\APACjournalVolNumPages{arXiv e-prints}{}{}{arXiv:2106.14550}.
\PrintBackRefs{\CurrentBib}

\bibitem [\protect \citeauthoryear {%
{France}%
\ \protect \BOthers {.}}{%
{France}%
\ \protect \BOthers {.}}{%
{\protect \APACyear {2013}}%
}]{%
France2013}
\APACinsertmetastar {%
France2013}%
\begin{APACrefauthors}%
{France}, K.%
, {Froning}, C\BPBI S.%
, {Linsky}, J\BPBI L.%
\ et al.\end{APACrefauthors}%
\unskip\
\newblock
\APACrefYearMonthDay{2013}{{\APACmonth{02}}}{},
\newblock
\unskip
\newblock
\APACjournalVolNumPages{\apj}{763}{2}{149}.
\newblock
\begin{APACrefDOI} \doi{10.1088/0004-637X/763/2/149} \end{APACrefDOI}
\PrintBackRefs{\CurrentBib}

\bibitem [\protect \citeauthoryear {%
{Go{\'z}dziewski}%
\ \protect \BOthers {.}}{%
{Go{\'z}dziewski}%
\ \protect \BOthers {.}}{%
{\protect \APACyear {2015}}%
}]{%
Gozdziewski2015}
\APACinsertmetastar {%
Gozdziewski2015}%
\begin{APACrefauthors}%
{Go{\'z}dziewski}, K.%
, {S{\l}owikowska}, A.%
, {Dimitrov}, D.%
\ et al.\end{APACrefauthors}%
\unskip\
\newblock
\APACrefYearMonthDay{2015}{{\APACmonth{04}}}{},
\newblock
\unskip
\newblock
\APACjournalVolNumPages{\mnras}{448}{2}{1118-1136}.
\newblock
\begin{APACrefDOI} \doi{10.1093/mnras/stu2728} \end{APACrefDOI}
\PrintBackRefs{\CurrentBib}

\bibitem [\protect \citeauthoryear {%
{Lecavelier Des Etangs}%
\ \protect \BOthers {.}}{%
{Lecavelier Des Etangs}%
\ \protect \BOthers {.}}{%
{\protect \APACyear {2010}}%
}]{%
Lecavelier2010}
\APACinsertmetastar {%
Lecavelier2010}%
\begin{APACrefauthors}%
{Lecavelier Des Etangs}, A.%
, {Ehrenreich}, D.%
, {Vidal-Madjar}, A.%
\ et al.\end{APACrefauthors}%
\unskip\
\newblock
\APACrefYearMonthDay{2010}{{\APACmonth{05}}}{},
\newblock
\unskip
\newblock
\APACjournalVolNumPages{\aap}{514}{}{A72}.
\newblock
\begin{APACrefDOI} \doi{10.1051/0004-6361/200913347} \end{APACrefDOI}
\PrintBackRefs{\CurrentBib}

\bibitem [\protect \citeauthoryear {%
{Lopez}%
, {Fortney}%
\BCBL {}\ \BBA {} {Miller}%
}{%
{Lopez}%
\ \protect \BOthers {.}}{%
{\protect \APACyear {2012}}%
}]{%
Lopez2012}
\APACinsertmetastar {%
Lopez2012}%
\begin{APACrefauthors}%
{Lopez}, E\BPBI D.%
, {Fortney}, J\BPBI J.%
\BCBL {}\ \BBA {} {Miller}, N.%
\end{APACrefauthors}%
\unskip\
\newblock
\APACrefYearMonthDay{2012}{{\APACmonth{12}}}{},
\newblock
\unskip
\newblock
\APACjournalVolNumPages{\apj}{761}{1}{59}.
\newblock
\begin{APACrefDOI} \doi{10.1088/0004-637X/761/1/59} \end{APACrefDOI}
\PrintBackRefs{\CurrentBib}

\bibitem [\protect \citeauthoryear {%
{Mayor}%
\ \BBA {} {Queloz}%
}{%
{Mayor}%
\ \BBA {} {Queloz}%
}{%
{\protect \APACyear {1995}}%
}]{%
Mayor1995}
\APACinsertmetastar {%
Mayor1995}%
\begin{APACrefauthors}%
{Mayor}, M.%
\BCBT {}\ \BBA {} {Queloz}, D.%
\end{APACrefauthors}%
\unskip\
\newblock
\APACrefYearMonthDay{1995}{{\APACmonth{11}}}{},
\newblock
\unskip
\newblock
\APACjournalVolNumPages{\nat}{378}{6555}{355-359}.
\newblock
\begin{APACrefDOI} \doi{10.1038/378355a0} \end{APACrefDOI}
\PrintBackRefs{\CurrentBib}

\bibitem [\protect \citeauthoryear {%
{Mulders}%
, {Pascucci}%
, {Apai}%
\BCBL {}\ \BBA {} {Ciesla}%
}{%
{Mulders}%
\ \protect \BOthers {.}}{%
{\protect \APACyear {2018}}%
}]{%
Mulders2018}
\APACinsertmetastar {%
Mulders2018}%
\begin{APACrefauthors}%
{Mulders}, G\BPBI D.%
, {Pascucci}, I.%
, {Apai}, D.%
\BCBL {}\ \BBA {} {Ciesla}, F\BPBI J.%
\end{APACrefauthors}%
\unskip\
\newblock
\APACrefYearMonthDay{2018}{{\APACmonth{07}}}{},
\newblock
\unskip
\newblock
\APACjournalVolNumPages{\aj}{156}{1}{24}.
\newblock
\begin{APACrefDOI} \doi{10.3847/1538-3881/aac5ea} \end{APACrefDOI}
\PrintBackRefs{\CurrentBib}

\bibitem [\protect \citeauthoryear {%
{Murray-Clay}%
, {Chiang}%
\BCBL {}\ \BBA {} {Murray}%
}{%
{Murray-Clay}%
\ \protect \BOthers {.}}{%
{\protect \APACyear {2009}}%
}]{%
Murray-Clay2009}
\APACinsertmetastar {%
Murray-Clay2009}%
\begin{APACrefauthors}%
{Murray-Clay}, R\BPBI A.%
, {Chiang}, E\BPBI I.%
\BCBL {}\ \BBA {} {Murray}, N.%
\end{APACrefauthors}%
\unskip\
\newblock
\APACrefYearMonthDay{2009}{{\APACmonth{03}}}{},
\newblock
\unskip
\newblock
\APACjournalVolNumPages{\apj}{693}{1}{23-42}.
\newblock
\begin{APACrefDOI} \doi{10.1088/0004-637X/693/1/23} \end{APACrefDOI}
\PrintBackRefs{\CurrentBib}

\bibitem [\protect \citeauthoryear {%
{Newton}%
\ \protect \BOthers {.}}{%
{Newton}%
\ \protect \BOthers {.}}{%
{\protect \APACyear {2019}}%
}]{%
Newton2019}
\APACinsertmetastar {%
Newton2019}%
\begin{APACrefauthors}%
{Newton}, E\BPBI R.%
, {Mann}, A\BPBI W.%
, {Tofflemire}, B\BPBI M.%
\ et al.\end{APACrefauthors}%
\unskip\
\newblock
\APACrefYearMonthDay{2019}{{\APACmonth{07}}}{},
\newblock
\unskip
\newblock
\APACjournalVolNumPages{\apjl}{880}{1}{L17}.
\newblock
\begin{APACrefDOI} \doi{10.3847/2041-8213/ab2988} \end{APACrefDOI}
\PrintBackRefs{\CurrentBib}

\bibitem [\protect \citeauthoryear {%
{{\"O}berg}%
\ \BBA {} {Wordsworth}%
}{%
{{\"O}berg}%
\ \BBA {} {Wordsworth}%
}{%
{\protect \APACyear {2019}}%
}]{%
Oberg2019}
\APACinsertmetastar {%
Oberg2019}%
\begin{APACrefauthors}%
{{\"O}berg}, K\BPBI I.%
\BCBT {}\ \BBA {} {Wordsworth}, R.%
\end{APACrefauthors}%
\unskip\
\newblock
\APACrefYearMonthDay{2019}{{\APACmonth{11}}}{},
\newblock
\unskip
\newblock
\APACjournalVolNumPages{\aj}{158}{5}{194}.
\newblock
\begin{APACrefDOI} \doi{10.3847/1538-3881/ab46a8} \end{APACrefDOI}
\PrintBackRefs{\CurrentBib}

\bibitem [\protect \citeauthoryear {%
{Owen}%
\ \BBA {} {Jackson}%
}{%
{Owen}%
\ \BBA {} {Jackson}%
}{%
{\protect \APACyear {2012}}%
}]{%
Owen2012}
\APACinsertmetastar {%
Owen2012}%
\begin{APACrefauthors}%
{Owen}, J\BPBI E.%
\BCBT {}\ \BBA {} {Jackson}, A\BPBI P.%
\end{APACrefauthors}%
\unskip\
\newblock
\APACrefYearMonthDay{2012}{{\APACmonth{10}}}{},
\newblock
\unskip
\newblock
\APACjournalVolNumPages{\mnras}{425}{4}{2931-2947}.
\newblock
\begin{APACrefDOI} \doi{10.1111/j.1365-2966.2012.21481.x} \end{APACrefDOI}
\PrintBackRefs{\CurrentBib}

\bibitem [\protect \citeauthoryear {%
{Poppenhaeger}%
, {Ketzer}%
\BCBL {}\ \BBA {} {Mallonn}%
}{%
{Poppenhaeger}%
\ \protect \BOthers {.}}{%
{\protect \APACyear {2021}}%
}]{%
Poppenhaeger2021}
\APACinsertmetastar {%
Poppenhaeger2021}%
\begin{APACrefauthors}%
{Poppenhaeger}, K.%
, {Ketzer}, L.%
\BCBL {}\ \BBA {} {Mallonn}, M.%
\end{APACrefauthors}%
\unskip\
\newblock
\APACrefYearMonthDay{2021}{{\APACmonth{01}}}{},
\newblock
\unskip
\newblock
\APACjournalVolNumPages{\mnras}{500}{4}{4560-4572}.
\newblock
\begin{APACrefDOI} \doi{10.1093/mnras/staa1462} \end{APACrefDOI}
\PrintBackRefs{\CurrentBib}

\bibitem [\protect \citeauthoryear {%
{Predehl}%
\ \protect \BOthers {.}}{%
{Predehl}%
\ \protect \BOthers {.}}{%
{\protect \APACyear {2021}}%
}]{%
Predehl2021}
\APACinsertmetastar {%
Predehl2021}%
\begin{APACrefauthors}%
{Predehl}, P.%
, {Andritschke}, R.%
, {Arefiev}, V.%
\ et al.\end{APACrefauthors}%
\unskip\
\newblock
\APACrefYearMonthDay{2021}{{\APACmonth{03}}}{},
\newblock
\unskip
\newblock
\APACjournalVolNumPages{\aap}{647}{}{A1}.
\newblock
\begin{APACrefDOI} \doi{10.1051/0004-6361/202039313} \end{APACrefDOI}
\PrintBackRefs{\CurrentBib}

\bibitem [\protect \citeauthoryear {%
{Rogers}%
}{%
{Rogers}%
}{%
{\protect \APACyear {2015}}%
}]{%
Rogers2015}
\APACinsertmetastar {%
Rogers2015}%
\begin{APACrefauthors}%
{Rogers}, L\BPBI A.%
\end{APACrefauthors}%
\unskip\
\newblock
\APACrefYearMonthDay{2015}{{\APACmonth{03}}}{},
\newblock
\unskip
\newblock
\APACjournalVolNumPages{\apj}{801}{1}{41}.
\newblock
\begin{APACrefDOI} \doi{10.1088/0004-637X/801/1/41} \end{APACrefDOI}
\PrintBackRefs{\CurrentBib}

\bibitem [\protect \citeauthoryear {%
{Salz}%
, {Schneider}%
, {Czesla}%
\BCBL {}\ \BBA {} {Schmitt}%
}{%
{Salz}%
\ \protect \BOthers {.}}{%
{\protect \APACyear {2016}}%
}]{%
Salz2016}
\APACinsertmetastar {%
Salz2016}%
\begin{APACrefauthors}%
{Salz}, M.%
, {Schneider}, P\BPBI C.%
, {Czesla}, S.%
\BCBL {}\ \BBA {} {Schmitt}, J\BPBI H\BPBI M\BPBI M.%
\end{APACrefauthors}%
\unskip\
\newblock
\APACrefYearMonthDay{2016}{{\APACmonth{01}}}{},
\newblock
\unskip
\newblock
\APACjournalVolNumPages{\aap}{585}{}{L2}.
\newblock
\begin{APACrefDOI} \doi{10.1051/0004-6361/201527042} \end{APACrefDOI}
\PrintBackRefs{\CurrentBib}

\bibitem [\protect \citeauthoryear {%
{Sanz-Forcada}%
\ \protect \BOthers {.}}{%
{Sanz-Forcada}%
\ \protect \BOthers {.}}{%
{\protect \APACyear {2011}}%
}]{%
Sanz-Forcada2011}
\APACinsertmetastar {%
Sanz-Forcada2011}%
\begin{APACrefauthors}%
{Sanz-Forcada}, J.%
, {Micela}, G.%
, {Ribas}, I.%
\ et al.\end{APACrefauthors}%
\unskip\
\newblock
\APACrefYearMonthDay{2011}{{\APACmonth{08}}}{},
\newblock
\unskip
\newblock
\APACjournalVolNumPages{\aap}{532}{}{A6}.
\newblock
\begin{APACrefDOI} \doi{10.1051/0004-6361/201116594} \end{APACrefDOI}
\PrintBackRefs{\CurrentBib}

\bibitem [\protect \citeauthoryear {%
{Schwope}%
\ \BBA {} {Thinius}%
}{%
{Schwope}%
\ \BBA {} {Thinius}%
}{%
{\protect \APACyear {2014}}%
}]{%
Schwope2014}
\APACinsertmetastar {%
Schwope2014}%
\begin{APACrefauthors}%
{Schwope}, A\BPBI D.%
\BCBT {}\ \BBA {} {Thinius}, B\BPBI D.%
\end{APACrefauthors}%
\unskip\
\newblock
\APACrefYearMonthDay{2014}{{\APACmonth{01}}}{},
\newblock
\unskip
\newblock
\APACjournalVolNumPages{Astronomische Nachrichten}{335}{4}{357}.
\newblock
\begin{APACrefDOI} \doi{10.1002/asna.201312053} \end{APACrefDOI}
\PrintBackRefs{\CurrentBib}

\bibitem [\protect \citeauthoryear {%
{Sunyaev}%
\ \protect \BOthers {.}}{%
{Sunyaev}%
\ \protect \BOthers {.}}{%
{\protect \APACyear {2021}}%
}]{%
Sunyaev2021}
\APACinsertmetastar {%
Sunyaev2021}%
\begin{APACrefauthors}%
{Sunyaev}, R.%
, {Arefiev}, V.%
, {Babyshkin}, V.%
\ et al.\end{APACrefauthors}%
\unskip\
\newblock
\APACrefYearMonthDay{2021}{{\APACmonth{04}}}{},
\newblock
\unskip
\newblock
\APACjournalVolNumPages{arXiv e-prints}{}{}{arXiv:2104.13267}.
\PrintBackRefs{\CurrentBib}

\bibitem [\protect \citeauthoryear {%
{Temple}%
\ \protect \BOthers {.}}{%
{Temple}%
\ \protect \BOthers {.}}{%
{\protect \APACyear {2019}}%
}]{%
Temple2019}
\APACinsertmetastar {%
Temple2019}%
\begin{APACrefauthors}%
{Temple}, L\BPBI Y.%
, {Hellier}, C.%
, {Anderson}, D\BPBI R.%
\ et al.\end{APACrefauthors}%
\unskip\
\newblock
\APACrefYearMonthDay{2019}{{\APACmonth{12}}}{},
\newblock
\unskip
\newblock
\APACjournalVolNumPages{\mnras}{490}{2}{2467-2474}.
\newblock
\begin{APACrefDOI} \doi{10.1093/mnras/stz2632} \end{APACrefDOI}
\PrintBackRefs{\CurrentBib}

\bibitem [\protect \citeauthoryear {%
{Watson}%
, {Donahue}%
\BCBL {}\ \BBA {} {Walker}%
}{%
{Watson}%
\ \protect \BOthers {.}}{%
{\protect \APACyear {1981}}%
}]{%
Watson1981}
\APACinsertmetastar {%
Watson1981}%
\begin{APACrefauthors}%
{Watson}, A\BPBI J.%
, {Donahue}, T\BPBI M.%
\BCBL {}\ \BBA {} {Walker}, J\BPBI C\BPBI G.%
\end{APACrefauthors}%
\unskip\
\newblock
\APACrefYearMonthDay{1981}{{\APACmonth{11}}}{},
\newblock
\unskip
\newblock
\APACjournalVolNumPages{\icarus}{48}{2}{150-166}.
\newblock
\begin{APACrefDOI} \doi{10.1016/0019-1035(81)90101-9} \end{APACrefDOI}
\PrintBackRefs{\CurrentBib}

\bibitem [\protect \citeauthoryear {%
{Wilms}%
, {Allen}%
\BCBL {}\ \BBA {} {McCray}%
}{%
{Wilms}%
\ \protect \BOthers {.}}{%
{\protect \APACyear {2000}}%
}]{%
Wilms2000}
\APACinsertmetastar {%
Wilms2000}%
\begin{APACrefauthors}%
{Wilms}, J.%
, {Allen}, A.%
\BCBL {}\ \BBA {} {McCray}, R.%
\end{APACrefauthors}%
\unskip\
\newblock
\APACrefYearMonthDay{2000}{{\APACmonth{10}}}{},
\newblock
\unskip
\newblock
\APACjournalVolNumPages{\apj}{542}{2}{914-924}.
\newblock
\begin{APACrefDOI} \doi{10.1086/317016} \end{APACrefDOI}
\PrintBackRefs{\CurrentBib}

\bibitem [\protect \citeauthoryear {%
{Yelle}%
}{%
{Yelle}%
}{%
{\protect \APACyear {2004}}%
}]{%
Yelle2004}
\APACinsertmetastar {%
Yelle2004}%
\begin{APACrefauthors}%
{Yelle}, R\BPBI V.%
\end{APACrefauthors}%
\unskip\
\newblock
\APACrefYearMonthDay{2004}{{\APACmonth{07}}}{},
\newblock
\unskip
\newblock
\APACjournalVolNumPages{\icarus}{170}{1}{167-179}.
\newblock
\begin{APACrefDOI} \doi{10.1016/j.icarus.2004.02.008} \end{APACrefDOI}
\PrintBackRefs{\CurrentBib}

\end{thebibliography}

\end{document}